\documentstyle[12pt]{article}
\begin{document}
\parskip=6pt
\baselineskip=20pt
{\raggedleft{ April 11, 1995\\}}
\bigskip
\centerline{\Large\bf Baxter-Bazhanov Model, Frenkel-Moore Equation}
\centerline{\Large\bf and the Braid Group}
\bigskip
\centerline{\large\bf Zhan-Ning Hu \footnote{\bf email address:
huzn@itp.ac.cn}}
\smallskip
\centerline{CCAST(WORLD LABORATORY) P.O.BOX 8730, BEIJING, 100080}
\centerline{and}
\centerline{INSTITUTE OF THEORETICAL PHYSICS, ACADEMIA SINICA,}
\centerline{ P. O. BOX 2735 BEIJING 100080, CHINA \footnote{\bf mail address}}
\vspace{2ex}
\bigskip
\begin{center}
\begin{minipage}{5in}
\centerline{\large\bf 	Abstract}
\vspace{1ex}
In this paper the three-dimensional vertex model is given, which is the
 duality of the three-dimensional Baxter-Bazhanov (BB) model. The  braid
group  corresponding to Frenkel-Moore equation is constructed and the
transformations $R, I$  are found. These maps act on the group and denote
 the rotations of the braids through the angles $\pi$ about some special
axes. The weight function of another three-dimensional vertex model related
 the 3D lattice integrable model proposed by Boos, Mangazeev, Sergeev and
Stroganov is presented also, which can be interpreted as the deformation
 of the vertex model corresponding to the BB model.

\vspace{5ex}
Short title: BB Model and Braid Group

\vspace{2ex}
0550, 0210, 0220, 0520
\end{minipage}
\end{center}
\newpage

\section{\bf Introduction}

The braid group has a deep connection with the Yang-Baxter equation which
plays an important role in the exactly solvable model in two dimensions in
statistical mechanics. The tetrahedron equation is an integrable condition of
statistical model in three-dimensions due to which gives the commutativity of
the layer to layer transfer matrixes  \cite{k2}. The Frenkel-Moore version of
the Tetrahedral equation  \cite{fm,cs}  is formulated as:
\begin{equation}
S_{123}S_{124}S_{134}S_{234}=S_{234}S_{134}S_{124}S_{123}
\end{equation}
where S$\in$End($V^{\otimes 3}$) and  each side of the equation acts on
$V^{\otimes 4}=V_1V_2V_3V_4$, and $S_{123}$, for example, acts on $V_4$
identically.
Then we can ask a question that what is the braid group
corresponding to Frenkel-Moore equation. An answer will be given in this paper.

 Just as the Yang-Baxter equation can describe the scattering  in two
dimensions,
 with a factorizable matrix, the tetrahedron equation gives the relations
among the scattering amplitudes of three strings \cite{z1}. And the three
labeling schemes exist \cite{jh}. With the cell (or vacuum) labeling Bazhanov
and Baxter generalized the two-state Zamolodchikov model \cite{z1} to  an
arbitrary number of  state \cite{b1}.  Then the three-dimensional star-star
relation is proved \cite{k1,h1,b2} and the star-square relation is discussed
\cite{k2,h2}. Recently Mangazeev and Boos $et  al$ obtained the solution of
 modified tetrahedron equation \cite{m2,bm} in terms of elliptic functions
which generalized the result of ref. \cite{m1}. Frenkel-Moore equation can
 also be gotten by using string labeling. So we discuss the braid relation of
 Frenkel-Moore equation in the following section by regarding the three-string
 cross as the ``elementary"  braid \cite{leehu}.

The vertex type tetrahedron equations were discussed in refs. \cite{Kor,jh} and
the discrete symmetry groups of vertex models were studied by Boukraa
$et~al$ \cite{bouk}. Mangazeev $et~al$ \cite{newsolu} proposed  a
three-dimensional vertex model and the weight function of this model can be
 obtained from Baxter-Bazhanov model \cite{humiwa} when taking some limits.
 The three-dimensional vertex model corresponding to Baxter-Bazhanov model is
 constructed in this paper. And the six spectrums  with a constrained
condition relate to the six spaces which the vertex type tetrahedron equation
is defined in.

This paper is organized as the follows. In section 2, the three-dimensional
 vertex model is given, which is a duality of the three-dimensional
 Baxter-Bazhanov model. The braid group corresponding to the Frenkel-Moore
 equation is constructed in section 3. In section 4, the transformations
 $R, I$  are discussed. Finally   some conclusions are given and Boltzmann
weight of another three-dimensional vertex model is presented. It corresponds
 to the 3D lattice integrable models proposed by Boos $et~al$ \cite{m2,bm}
 and can be regarded as the deformation of the vertex model corresponding
 to BB model.

\section{\bf The Three-Dimensional Vertex Model}

The Baxter-Bazhanov model is an Interaction-Round-a-Cube (IRC) Model. The
 weight function of it has the form \cite{k1}
$$
W_P(a|efg|bcd|h) ~~~~~~~~~~~~~~~~~~~~~~~~~~~~~~~~~~~~~~~~~~~~~~~~~~~~~~~~~
{}~~~~~~~
$$$$
=\frac{\omega^{fb}}{\omega^{ag}}\Bigg[\frac{w(x_{14}x_{23},x_{12}x_{34},
x_{13}x_{24}|a+d,e+f)}{w(x_{14}x_{23},x_{12}x_{34},x_{13}x_{24}|g+h,c+b)}
\Bigg]^{1/2}~~~
$$$$
\times \Bigg[\frac{w(x_4,x_{34},x_3|e+h,d+c)}{w(x_4,x_{34},x_3|a+b,f+g)}
\Bigg]^{1/2} ~~~~~~~~~~~~~~~~~~
$$$$
\times \Bigg[\frac{w(x_2,x_{12},x_1|e+g,a+c)}{w(x_2,x_{12},x_1|d+b,f+h)}
\Bigg]^{1/2}\frac{\omega^{(ag+gb+bh)/2}}{\omega^{(hd+de+ea)/2}}~
$$
\begin{equation} \label{W1}
{}~~~~~~~~~~~~~\times \Bigg\{\sum_{\sigma \in Z_N}
\frac{w(x_3,x_{13},x_1|d,h+\sigma)w(x_4,x_{24},x_2|a,g+\sigma)}
{w(x_4,x_{14},x_1|e,c+\sigma)w(x_3/\omega,x_{23},x_2|f,b+\sigma)}\Bigg\}_0.
\end{equation}
where the subscript ``0'' after the curly brackets indicates that the
 expression in the braces is divided by itself with the zero exterior
 spins and  we have used the notations
\begin{equation}
w(x,y,z|k,l)=w(x,y,z|k-l)\Phi(l),~~w(x,y,z|l)=\prod^{l}_{j=1}\frac{y}
{z-x\omega^j},~~k,l\in Z_N,
\end{equation}
\begin{equation}
x^N+y^N=z^N,~\Phi(l)=\omega^{l(l+N)/2},~\omega^{1/2}=\exp(\pi i/N),
{}~x^N_i-x^N_j=x^N_{ij},
\end{equation}
for $i<j$ and $i,j=1,2,3,4$.
It satisfies the cube type tetrahedron equation
$$
\sum_{d}
W(a_4|c_2c_1c_3|b_1b_3b_2|d)W'(c_1|b_2a_3b_1|c_4dc_6|b_4)  ~~~~~~~~~~
{}~~~~~~~~~~~~~~~~~~~~~
$$$$
\times W''(b_1|dc_4c_3|a_2b_3b_4|c_5)W'''(d|b_2b_4b_3|c_5c_2c_6|a_1)~~~~
{}~~~~~~~~~~~~~~~~~~~
$$$$
{}~~~~~~~~~~~~~~~~~=\sum_{d}
W'''(b_1|c_1c_4c_3|a_2a_4a_3|d)W''(c_1|b_2a_3a_4|dc_2c_6|a_1)
$$
\begin{equation} \label{cube}
{}~~~~~~~~~~~~~~~~~~~~~~~~~~~\times W'(a_4|c_2dc_3|a_2b_3a_1|c_5)
W(d|a_1a_3a_2|c_4c_5c_6|b_4)
\end{equation}
where $W$, $W'$, $W''$ and $W'''$ are some four
 sets of Boltzmann weights. By using the symmetry properties of the
Boltzmann weights and setting
\begin{equation}
u=\frac{x_1}{\omega x_2},~ v=\frac{x_4}{x_3},~z=\frac{z_1}{z_2},
{}~z_1=\frac{x_{13}}{\omega x_{14}},~z_2=\frac{x_{23}}{x_{24}},
\end{equation}
the  Boltzmann weight of the Baxter-Bazhanov model can be written
into the vertex form
$$
R(u,z,v)^{j_1j_2j_3}_{i_1i_2i_3}=(-)^{j_2}(\omega^{1/2})
^{j_1j_2+j_2j_3+j_1j_3}\Bigg[\frac{w(u,j_1)w(z_2/(\omega z_1),-i_2)w(v,i_3)}
{w(u,i_1)w(z_2/(\omega z_1),-j_2)w(v,j_3)}\Bigg]^{1/2}
$$
\begin{equation}
\times\Bigg\{\sum_{\sigma\in Z_N}\frac{w(\omega vz_1,\sigma+j_2+j_3)
w(z_2,\sigma)s(\sigma,j_1)}{w(z_1,\sigma+j_2)w(vz_2,\sigma+i_3)}\Bigg\}_0
\end{equation}
with
\begin{equation}
{w(v,a)\over w(v,0)}=[\Delta(v)]^a\prod^a_{j=1}(1-\omega^jv)^{-1},
{}~~\Delta(v)=(1-v^N)^{1/N}
\end{equation}
where $s(a,b)=\omega^{ab}$ and  the spin variables
$i_1$, $i_2$, $i_3$, $j_1$, $j_2$, $j_3$ satisfy
 the conditions $i_1+i_2=j_1+j_2,~ i_2+i_3=j_2+j_3$.  The Boltzmann weights
 satisfy the vertex type tetrahedron equation
$$
{\displaystyle\sum_{\{k_i\},\atop i=1,\cdots,6}}
R(u_1,u_2,u_3)^{k_1,k_2,k_3}_{i_1,i_2,i_3}
R(u_1,u_4,u_5)^{j_1k_4k_5}_{k_1i_4i_5}
R(u_2,u_4,u_6)^{j_2j_4k_6}_{k_2k_4i_6}
R(u_3,u_5,u_6)^{j_3j_5j_6}_{k_3k_5k_6}= ~~~~~~~
$$
\begin{equation}
{}~{\displaystyle\sum_{\{k_i\},\atop i=1,\cdots,6}}
R(u_3,u_5,u_6)^{k_3,k_5,k_6}_{i_3,i_5,i_6}
R(u_2,u_4,u_6)^{k_2k_4j_6}_{i_2i_4k_6}
R(u_1,u_4,u_5)^{k_1j_4j_5}_{i_1k_4k_5}
R(u_1,u_2,u_3)^{j_1j_2j_3}_{k_1k_2k_3}
\end{equation}
where
$$
u_1=\frac{x_1}{\omega x_2}=\frac{x_1'}{\omega x_2'},~~~~~~~~
{}~~~~u_2=\frac{x_{13}x_{24}}{\omega x_{14}x_{23}}=\frac{x_1''}
{\omega x_2''},~~u_3=\frac{x_4}{x_3}=\frac{x_1'''}{\omega x_2'''},
$$
\begin{equation}
u_4=\frac{x_{13}'x_{24}'}{\omega x_{14}'x_{23}'}=\frac{x_{13}''x_{24}''}
{\omega x_{14}''x_{23}''},~~u_5=\frac{x_4'}{x_3'}=\frac{x_{13}'''x_{24}'''}
{\omega x_{14}'''x_{23}'''},~~~~u_6=\frac{x_4''}{x_3''}=\frac{x_4'''}
{x_3'''}.~
\end{equation}
In this way, we get a three-dimensional vertex model \cite{hujsp} which
 corresponds to the Baxter-Bazhanov model. The spectrums $u_i
(i=1,2,\cdots,6)$ appeared in the above tetrahedron equation satisfy
 the condition
$$
\bigg[sin\frac{\theta_1+\theta_2+\theta_3}{2}
sin\frac{-\theta_1+\theta_2+\theta_3}{2}
sin\frac{-\theta_3+\theta_5+\theta_6}{2}
sin\frac{\theta_3+\theta_5-\theta_6}{2}\bigg]^{1/2}~~~~~~~~~~~~
$$$$
-\bigg[sin\frac{\theta_1-\theta_2+\theta_3}{2}
sin\frac{\theta_1+\theta_2-\theta_3}{2}
sin\frac{\theta_3-\theta_5+\theta_6}{2}
sin\frac{\theta_3+\theta_5+\theta_6}{2}\bigg]^{1/2}~~~~~
$$
\begin{equation}\label{3.2}
{}~~~~~~~~~~~~~~~~~~~~~~~~~~~~~~~~~~~~~~=sin\theta_3
\bigg[sin\frac{\theta_2+\theta_4-\theta_6}{2}
sin\frac{-\theta_2+\theta_4+\theta_6}{2}\bigg]^{1/2}
\end{equation}
where we have parametrized the spectrums of the Boltzmann
 weights as
\begin{equation}
u_i=\omega^{-1/2}[ctg(\frac{\theta_i}{2})]^{2/N}, ~~~~~~i=1,2,\cdots,6.
\end{equation}

\section{\bf Braid Group \^B$_N$}

The shorthand notation of the vertex type tetrahedron equation (9) is
\begin{equation}
R_{123}R_{145}R_{246}R_{356}=R_{356}R_{246}R_{145}R_{123},
\end{equation}
which can be reformulated as \cite{Kor}
\begin{equation}
R_{12,13,23}R_{12,14,24}R_{13,14,34}R_{23,24,34}=R_{23,24,34}R_{13,14,34}
R_{12,14,24}R_{12,13,23}
\end{equation}
by using the following translation \cite{jh}:
\begin{equation}
1\rightarrow 12, ~~2\rightarrow 13,~~3\rightarrow 23,~4\rightarrow 14,
{}~~5\rightarrow 24,~~6\rightarrow 34.
\end{equation}
Then we can  wright down the Frenkel-Moore equation (1) (see ref.\cite{jh}).
The relations of the braid group corresponding to the Frenkel-Moore equation
are discussed as follows.  For $N+1$ strings, let us express the
``elementary" braid $\alpha_i$  and
$ \alpha^{-1}_i  (i=1, 2, \cdots, N-1)$  by Fig.1 and Fig.2 respectively.
 In
 Fig.1, $i+1$-string is on the bottom, $i-1$-string is on the top and
$i$-string is between the $i-1$-string and $i+1$-string. In Fig.2,
 $i-1$-string is on the bottom, $i+1$-string is on the top and $i$-string
 is
also between the $i-1$-string and $i+1$-string.  When we define the product
 of
$\alpha_i,$ and  $\alpha_j  (1 \leq i,j \leq N-1)$ as that $\alpha_j$ acts on
$N+1$ strings on which  $\alpha_i$ has acted, $\alpha_i, \alpha^{-1}_i,
\alpha_j,  \alpha^{-1}_j,  \cdots $ and all of the arbitrary products of them
 form a group  \^B$_N$ in which the identity element is the no crossed  $N+1$
strings. The ``elementary" braid $\alpha_i$  satisfy the following relations:
 \begin{equation}  \begin{array}{cc}
\alpha_i \alpha_j = \alpha_j \alpha_i ,  &  |i-j|\geq 3 , \\
 \end{array}
 \end{equation}
 \begin{equation}
\alpha_i \alpha_{i\pm 1} \alpha_i \alpha_{i\pm 1}
=\alpha_{i\pm 1} \alpha_i \alpha_{i\pm 1} \alpha_i,
 \end{equation}
 \begin{equation}
\alpha_{i\pm 1} \alpha_{i\mp 1} \alpha_{i\pm 1} \alpha_i
=\alpha_i \alpha_{i\mp 1} \alpha_{i\pm 1} \alpha_{i\mp 1},
 \end{equation}
 \begin{equation}
\alpha_{i\pm 1} \alpha_{i\mp 1} \alpha_i \alpha_{i\pm 1}
= \alpha_{i\mp 1}\alpha_i \alpha_{i\pm 1} \alpha_{i\mp 1},
 \end{equation}
 \begin{equation}
\alpha_i \alpha ^{-1}_i = \alpha ^{-1}_i \alpha_i =E   .
 \end{equation}
Notice that there are two relations in equations (17), (18) and (19),
respectively. The first ones of them can be proved graphically by Fig.3,
 Fig.4 and
Fig.5. The second relations of them can be proved similarly. From these
 relations it can be gotten easily that
 \begin{equation}
\alpha_{i\pm 1} \alpha_{i\mp 1} (\alpha_{i\pm 1} \alpha_i)^{2n+1}
=(\alpha_i \alpha_{i\mp 1})^{2n+1} \alpha_{i\pm 1} \alpha_{i\mp 1},
 \end{equation}
 \begin{equation}
\alpha_{i\pm 1} \alpha_{i\mp 1} (\alpha_i  \alpha_{i\pm 1})^{2m+1}
=(\alpha_{i\mp 1} \alpha_i)^{2m+1} \alpha_{i\pm 1} \alpha_{i\mp 1},
 \end{equation}
 \begin{equation}
\alpha_{i\pm 1} \alpha_{i\mp 1} (\alpha_i  \alpha_{i\pm 1})^{2l}
=(\alpha_i \alpha_{i\mp 1})^{2l} \alpha_{i\pm 1} \alpha_{i\mp 1},
 \end{equation}
where $m, n, l,$ are all the arbitrary integers.

Now we discuss the generators of  the braid group \^B$_N  (N\geq 2)$  as
 follows. Firstly ,  \^B$_2 $  and  \^B$_3$ have two and four
 generators respectively. And \^B$_4 $ has also four generators owing to
 $\alpha_2 =\alpha^{-1}_1 \beta^2 \alpha^{-1}_1 \beta^{-1}$ and $
  \alpha_3 = \beta \alpha_1 \beta^{-1}$ where  $\beta = \alpha_1 \alpha_2
 \alpha_3$. For
$N\geq 5$, setting
\begin{equation}
\beta=\alpha_1 \alpha_2 \alpha_3 \cdots \alpha_{N-1} ,
\end {equation}
we have, from equation (19),
\begin{equation}  \begin{array}{cc}
\alpha_i \beta=\beta \alpha_{i-2} , &  N-1 \geq i \geq 3. \\
\end{array}
\end {equation}
Then  $\alpha_i $ can be expressed as
 \begin{equation}
\alpha_i=\beta^{[\frac{i-1}{2}]} \alpha_{i-2[\frac{i-1}{2}]}
\beta^{-[\frac{i-1}{2}]}
 \end{equation}
where $[\frac{i-1}{2}]$ is an integer but  $\frac{i}{2}-1 \leq [\frac{i-1}{2}]
\leq \frac{i-1}{2}$.  So braid group \^B$_N (N\geq 5)$   has six generators:
 $\alpha_1, \alpha_2, \beta$ and the inverses of them: $\alpha^{
-1}_1, \alpha^{-1}_2$ and  $\beta^{-1}$ due to
 \begin{equation}
\alpha_{i-2[\frac{i-1}{2}]}= \left\{
\begin{array}{ll}
\alpha_1  & \mbox{for odd $i$}\\
\alpha_2 & \mbox{for even $i$}
\end{array}.
\right.
 \end{equation}
The equation (17), that is, Fig.3, is the braid relation for the
Frenkel-Moore equation which been written as \cite{leehu}
\begin{equation} \label{eq1}
S_{123}S_{214}S_{143}S_{234}=S_{234}S_{143}S_{214}S_{123}.
\end{equation}
When we denote the indexes of $S$  by the place where the strings cross, we
have\begin{equation} \label{eq2}
S_{123}S_{234}S_{123}S_{234}=S_{234}S_{123}S_{234}S_{123}.
\end{equation}
This is just the planar tetrahedral equation\cite{cs}.

\section{\bf Transformations $R$  and  $I$}

{}From relations (18) and (19) we get
\begin{equation} \label{eq3}
S_{123}S_{145}S_{325}S_{234}=S_{234}S_{325}S_{145}S_{123},
\end{equation}
\begin{equation} \label{eq4}
S_{123}S_{145}S_{254}S_{345}=S_{345}S_{254}S_{145}S_{123}.
\end{equation}
The other two equations corresponding to relations (18) and (19)  can be
 obtained from the above relations by using the index maps:
$1 \longleftrightarrow 5$ and $2 \longleftrightarrow  4$. This means that
 some relations exist between the two equations in (17) or in (18),
respectively. So we set
\begin{equation}  \begin{array}{lc}
R(\alpha_i)=\alpha_{\sigma (i)}, & R(\alpha_i \alpha_j)=\alpha_{\sigma (i)}
\alpha_{\sigma (j)},\\
R(\alpha_i \alpha_j \alpha_k)=\alpha_{\sigma (i)} \alpha_{\sigma (j)}
 \alpha_{\sigma (k)}, &  \cdots ;
\end{array}
\end{equation}
\begin{equation} \begin{array}{lc}
I(\alpha_i)=\alpha_i, & I(\alpha_i \alpha_j)=\alpha_j \alpha_i, \\
I(\alpha_i \alpha_j \alpha_k)=\alpha_k \alpha_j \alpha_i , \; \; \; \; \; \; \;
\; \;  \; \; \; \; \; \; \; \;\; \; &  \cdots .
\end{array} \end{equation}
where $ \sigma (i)=N-i$ and $N$ is the total numbers of strings minus one.
It can be proved easily that
\begin{equation} \begin{array}{cc}
[R,I]=0, &  R^2 = I^2 =1. \\
\end{array} \end{equation}
The transformation $R$ acting  on  the braids denotes the rotation of the
 braids through the angle $\pi $ around the axis  which the strings cross
along. The transformation $I$ acting on the braids denotes that the braids are
rotated through  the angle $\pi$ about the axis  which is in the plan which
the braids are in and is perpendicular to the  above axis. Then transformation
 $RI$  describes the rotation of the braids through the angle $\pi $ for the
axis which is perpendicular to the plan in which the braids are. So the two
relations in (18)  or (19) can be changed each other by using the
transformation $I$ and (then) we can choice only one of them respectively.
Acting on the generators: $\Delta =b_1 b_2 \cdots b_{N-1}$ and  $\Omega =b_1
\Delta$, of the ordinary braid group $B_N$ by the transformations $R$ and $I$,
 we have that
\begin{equation} \begin{array}{l}
R(\Delta )= I(\Delta )=\Delta^{N-1} (\Delta^{-1} \Omega \Delta^{-1} )^{N-1}, \\
R(\Omega )=\Delta^{N-2} \Omega (\Delta^{-1} \Omega \Delta^{-1} )^{N-1}, \\
I(\Omega )= \Delta^{N-1} (\Delta^{-1} \Omega \Delta^{-1} )^{N-1} \Omega
\Delta^{-1}.
\end{array} \end{equation}
It is note that the braid relations of the planar permutohedron equation
can
be gotten easily from elements of $B_N$ and \^B$_N$.

\section{\bf Conclusions and Remarks}

Similarly as the above discussion, we can wright down the weight function of
 the three-dimensional vertex model related the 3D lattice integrable model
 proposed
 in refs. \cite{m2,bm} as the following form:
$$
R(u_1,u_2,u_3)^{j_1j_2j_3}_{i_1i_2i_3}=(-)^{j_2}(\omega^{1/2})^
{j_1j_2+j_2j_3+j_1j_3}~~~~~~~~~~~~~~~~~~~~~
$$$$
{}~~~~~~~~~~~~~~~~~\times\Bigg[\frac{w(qu_1,j_1)w(q^{-1}(\omega u_2)^{-1},-i_2)
w(q^{-1}u_3,i_3)}{w(q^{-1}u_1,i_1)w(q(\omega u_2)^{-1},-j_2)w(qu_3,j_3)}
\Bigg]^{1/2}\times
$$
\begin{equation}
\times\Bigg\{\sum_{\sigma\in Z_N}\frac{w(\omega u_2'u_3,\sigma+j_2+j_3)
w(qu_2'',\sigma)s(\sigma,j_1)}{w(q^{-1}u_2',\sigma+j_2)w(u_2''u_3,
\sigma+i_3)}\Bigg\}_0
\end{equation}
It satisfies the modified tetrahedron equation
\begin{equation}
R_{123}\overline{R}_{145}R_{246}\overline{R}_{356}=R_{356}
\overline{R}_{246}R_{145}\overline{R}_{123},
\end{equation}
where $\overline{R}(u_1,u_2,u_3)^{j_1j_2j_3}_{i_1i_2i_3}$ can be
 obtained from expression (36) by the substitutions:
\begin{equation}
q\rightarrow q^{-1},~~u_2'\rightarrow \overline{u}_2',~~u_2''\rightarrow
 \overline{u}_2'',
\end{equation}
with the condition $u_2'\overline{u}_2''=u_2''\overline{u}_2'$. The details
 will be given else where. When $q=1$, this vertex model reduces to the one
in section 2. Then it can be regarded as a deformation of the
 three-dimensional vertex model corresponding to BB model.

As the conclusions we get the three-dimensional vertex model which is a
duality of the three-dimensional Baxter-Bazhanov model. The Boltzmann weights
 of the model satisfy the vertex type tetrahedron equation. The braid group
 \^B$_N$ corresponding to the
Frenkel-Moore equation is constructed and the transformations $R$ and $I$
 acting on
 the braids and denoting the rotations of the braids around some special
 axes
through the angle $\pi$ are given. Using the method presented in this paper,
 we can constructed a new three-dimensional vertex model for which the weight
 function has the form (36). It is a duality of the 3D lattice integrable
model proposed in refs. \cite{m2,bm} and a deformation of the vertex model
related BB model. This means that we can interpreted the 3D lattice
integrable model in ref. \cite{m2,bm} as a deformation of the
three-dimensional Baxter-Bazhanov model. The symmetry properties of the
weight function (36) can be given similar as in ref. \cite{hujsp}.

\section{\bf Acknowledgment}

The author would like to thank H. Y. Guo, B. Y. Hou, K. J. Shi, P. Wang
and K. Wu for the interesting discussions and C. S. Huang, Z. C. Ou-Yang
 for the encouragements in this work.

\bigskip
\bigskip

\newpage

\section*{\bf Captions}

\* \* \* \* \* 1.  Fig.1  $\alpha_i$

2.  Fig.2  $\alpha^{-1}_i$

3.  Fig.3  the graphic proof of $\alpha_i \alpha_{i+1} \alpha_i \alpha_{i+1}
=\alpha_{i+1} \alpha_i \alpha_{i+1} \alpha_i $

4.  Fig.4  the graphic proof of $ \alpha_{i+1} \alpha_{i-1} \alpha_{i+1}
 \alpha_i =\alpha_i \alpha_{i-1} \alpha_{i+1} \alpha_{i-1} $

5.  Fig.5  the graphic proof of $ \alpha_{i+1} \alpha_{i-1} \alpha_i
\alpha_{i+1} = \alpha_{i-1}\alpha_i \alpha_{i+1} \alpha_{i-1} $

\newpage
{}~~

\vspace{7cm}
\hspace{2cm} Fig.1 $\alpha_i$  \hspace{6cm} Fig.2  $\alpha^{-1}_i$

\smallskip
\vbox{\hbox{Author: Zhan-Ning Hu ~~~~~~~~Title:  BB Model and Braid Group}}

\vspace{9cm}
\centerline{ Fig.3  the graphic proof of $\alpha_i \alpha_{i+1} \alpha_i
\alpha_{i+1}=\alpha_{i+1} \alpha_i \alpha_{i+1} \alpha_i $}

\smallskip
\vbox{\hbox{Author: Zhan-Ning Hu ~~~~~~~~Title:  BB Model and Braid Group}}

\newpage
{}~~

\vspace{7cm}
\centerline{ Fig.4  the graphic proof of $ \alpha_{i+1} \alpha_{i-1}
\alpha_{i+1} \alpha_i =\alpha_i \alpha_{i-1} \alpha_{i+1} \alpha_{i-1} $}

\smallskip
\vbox{\hbox{Author: Zhan-Ning Hu ~~~~~~~~Title:  BB Model and Braid Group}}

\vspace{9cm}
\centerline{ Fig.5  the graphic proof of $ \alpha_{i+1} \alpha_{i-1} \alpha_i
\alpha_{i+1} = \alpha_{i-1}\alpha_i \alpha_{i+1} \alpha_{i-1} $}

\smallskip
\vbox{\hbox{Author: Zhan-Ning Hu ~~~~~~~~Title:  BB Model and Braid Group}}

\end{document}